\documentclass[useAMS,usenatbib,usegraphicx]{mn2e}
\usepackage{amsmath}
\usepackage{txfonts}
\usepackage{float}
\usepackage{subfigure}
\usepackage[T1]{fontenc}
\usepackage{aecompl}
\usepackage{longtable}\DeclareMathAlphabet{\mathpzc}{OT1}{pzc}{m}{it}
\usepackage{color}
\font\sc=cmcsc10
\input epsf
\usepackage[authoryear]{natbib}
\newcommand{\beq}{\begin{equation}}
\newcommand{\eeq}{\end{equation}}
\newcommand{\noi}{\noindent}
\newcommand{\bneq}{\begin{eqnarray}}
\newcommand{\eneq}{\end{eqnarray}}
\newcommand{\bet}{\begin{table}}
\newcommand{\et}{\end{table}}
\newcommand{\btab}{\begin{tabular}}
\newcommand{\etab}{\end{tabular}}

\title[Chemical abundances and the evolution of Ursa Minor]{An inefficient dwarf: Chemical abundances and the evolution of the Ursa Minor dwarf spheroidal galaxy}

\author[U. Ural, et al.]
{U\u{g}ur Ural$^1$,
Gabriele Cescutti$^1$,
Andreas Koch$^{2}$, 
Jan Kleyna$^3$,
Sofia Feltzing$^4$, \newauthor 
Mark I. Wilkinson$^5$ \\\\
$^1$Leibniz Institut f\"{u}r Astrophysik Potsdam (AIP), An der Sternwarte 16, 14482 Postdam, Germany\\
$^2$Landessternwarte, Zentrum f\"ur Astronomie der Universit\"at Heidelberg, K\"onigstuhl 12, 69117 Heidelberg, Germany\\
$^3$Institute for Astronomy (IfA), University of Hawaii, 2680 Woodlawn Drive, Honolulu, HI, 96822, USA\\
$^4$Lund Observatory, Department of Astronomy and Theoretical Physics, Box 43, 22100 Lund, Sweden.\\
$^5$Department of Physics and Astronomy, University of Leicester, University Road, Leicester, LE1 7RH, United Kingdom .\\
Accepted for publication in MNRAS\\
}

\begin{document} 

\maketitle

\begin{abstract} 

\noi We present detailed chemical element abundance ratios of 17 elements with 8$\leq$~Z~$\leq$60 in three 
metal poor stars in the Ursa Minor dwarf spheroidal galaxy, which we combine with extant data from the literature to assess the predictions of a novel suite of galaxy chemical evolution models. The spectroscopic 
data were obtained with the Keck/HIRES instrument and revealed low metallicities
 of [Fe$/$H]=$-$2.12, $-$2.13 and $-$2.67~dex. While the most metal poor star in our
 sample shows an overabundance of [Mn/Fe] and other Fe-peak elements, our overall findings
 are in agreement with previous studies of this galaxy: elevated
 values of the [$\alpha$/Fe] ratios that are similar to, or only slightly lower than, the  
halo values but with SN~Ia enrichment at very low metallicity, as well as an enhancement 
of the ratio of first to second peak 
neutron capture elements [Y/Ba] with decreasing metallicity. The chemical evolution models which were tailored 
to reproduce the metallicity distribution function of the dSph,  indicate that UMi had an 
 extended star formation which lasted nearly 5~Gyr with low efficiency and are able to explain the
[Y/Ba] enhancement at low metallicity for the first time. In particular, we
show that the present day lack of gas is probably due to continuous loss of gas from the system, which we model as winds.

\end{abstract}


\section{Introduction}
\label{sec:intro}

\noi The increasing number of detailed spectroscopic surveys investigating 
kinematic and chemical properties of dwarf spheroidal (dSph) galaxies \citep[see e.g.,][]{Tolstoy2004,Helmi2006,Koch2006,Martin2007,Walker2009a} provide
new opportunities to study their local formation environment in detail and the formation 
and evolution of galaxies in general. 
Despite their low luminosities, their gravitational potentials had to be large 
enough to confine their gas for sufficient time to allow an extended period of star 
formation (SF) to occur. This is indicated by the large spreads in their metallicities and by 
the presence of stellar populations as young as 1-2 Gyr \citep[and references therein]{Tolstoy2009},
even down to 200 Myr \citep{Stetson1998,deBoer2012}.
 Samples of even a few stars in faint dSphs can provide interesting information \citep{Koch2008, Simon2010} as they emphasise the diversity of the chemical inventories of individual dSphs and guide us through the processes that affected their evolution~\citep{Tolstoy2009}. 

\noi One of the most important observed signatures is the spread in the ratio of 
[$\alpha/$Fe] in dSphs where the downturn of [$\alpha/$Fe] with increasing metallicity marks the time scale on which SN~Ia started to contribute to the enrichment of the interstellar medium (ISM). Other mechanisms such as episodic SF~\citep{Hendricks2014} or a change in the Initial Mass Function (IMF) where the very massive stars do not occur~\citep{McWilliam2013} might also be responsible for the spread found in the abundances of the $\alpha$-elements in the present day dSphs.

\begin{table*}
\begin{center}
\begin{tabular}[!hb]{l c c c c c c c c c c r}
\hline
Target &  I & V & K &$\alpha$ {J2000} & $\delta$ (J2000) & $\rm T_{eff}$ (phot) & $\rm T_{eff}$ (spec) &  log \it g & $\rm v_{mic}$ & [Fe$/$H]& $v_{\rm r}$ (km\,s$^{-1}$)  \\\hline
 UMI718  &16.30 & 17.46 & 14.967 & 15:02:25.58 & 66:55:25.39 & 4563 &  4540 & 1.1 & 2.25 & -2.12 & -247.6 \\
 UMI396  & 15.59 & 16.94 & 14.018 & 15:05:52.04 & 66:41:30.05 & 4312 & 4290 & 0.5 & 2.35 & -2.13 & -228\\
 UMI446  &  16.96 & 18.07 & 15.131 & 15:02:01.53 & 66:49:22.36 & 4640 & 4580 & 1.6 & 2.80 &-2.67 &-246 \\
\end{tabular}
\caption{ Observed stars: The K magnitudes are obtained from 2MASS photometry while I and V are from \protect\cite{Kleyna1998}. The preliminary estimates for the photometric temperature T(phot) from (V$-$I) and the surface gravity are calculated assuming a metallicity of [Fe$/$H]$=-$2 dex, mass of 0.8M$_\odot$ and distance of 68~kpc.The T(spec), log g, $\rm v_{mic}$ and [Fe/H] are the values found by iteration of the best atmospheric models used to calculate the spectroscopic abundances (see text for details). }
\end{center}
\label{tab:obs}
\end{table*}

\noi  Although many isolated dwarf galaxies seem to be dominated by exclusively old and/or young stellar populations~\citep{Benitez2014}, among the dSphs, Ursa Minor (UMi) is a rare system which has exclusively old stars and hence a candidate for being a dSph similar to those which contributed to the growth of the outer halo. \cite{Carrera2002} suggested that 95\% of its stars are older than 10~Gyrs and \cite{Ikuta2002} constrain the duration of SF in UMi to be between 3.9 to 6.5~Gyr. In this paper, we add  the chemical abundances of three red giant branch stars in UMi to the eighteen stars from the  previous studies by \cite{Cohen2010, Shetrone2001, Sadakane2004, Kirby2012}.  The chemical evolution models tested by \cite{Kirby2011} and \cite{Kirby2013} showed that the metallicity distribution function (MDF) of UMi can only be explained with models without extended infall of gas. We present detailed chemical evolution models describing the nucleosynthesis of 12 elements for the combination of our new data set with the extant data, examine these findings and study the implications for individual elements.
 
\noi The paper is organised as follows: In Section~\ref{sec:umiobs}, both the observations and the methods used to obtain 
the chemical abundances are explained and the final abundances for $\alpha$, Fe-peak 
and neutron capture elements are given for three new stars. In Section~\ref{sec:model}, 
we explain the details of our chemical evolution models
 and compare them to the combined data set. In Section~\ref{sec:disc}, we 
discuss our results and open questions. 

\section{Observations and Analysis}
\label{sec:umiobs}

\subsection{Observations and data reduction}
\label{subsec:reduction}

\noi The observations were made using the High Resolution Echelle Spectrograph (HIRES; \cite{Vogt1994}) on the Keck telescope
on the night of UT 2006-06-05. The targets were selected from legacy KPNO 4~m
MOSAIC imaging as in \cite{Wilkinson2004}. Details of the targets are given in Table~\ref{tab:obs}, together with some of their basic characteristics. 

\noi We used HIRES configured with the red collimator,  the B2 decker set for a fully open slit, the kv408 filter with the cut off at 450~nm, an echelle angle of 0◦.11197 and cross-disperser angle of 0◦.65899. We obtained a full wavelength
coverage of 440-750~nm with a 10~nm wide gap at 590~nm. The exposure times were 4000s for UMI718, 3600s for UMI396 and 3000s for UMI446.

\noi The raw data were reduced using the MAKEE data reduction package\footnote{MAKEE was developed by T. A. Barlow specifically for reduction of Keck HIRES data. 
It is freely available on the World Wide Web at the Keck Observatory home page, \tt{http://www2.keck.hawaii.edu/inst/hires/makeewww}}, which performs standard steps such as bias and flat field corrections, wavelength calibration via Th-Ar lamps, 
optimal subtraction, and sky subtraction. This procedure resulted in  signal-to-noise (S/N) ratios measured in the H$_{\alpha}$ order around 660~nm of 25, 30, 12 per pixel, respectively, for UMI718, UMI396 and UMI446. 

\noi The continuum fit to each echelle order was obtained by dividing  by
high-order polynomials within the {\tt continuum} task in IRAF\footnote{IRAF is distributed by the National Optical Astronomy Observatory, which is operated by the Association of Universities for Research in Astronomy (AURA) under cooperative agreement with the National Science Foundation}. Subsequently, 
 the velocity correction from vacuum to air velocities was made using 
the {\tt disptrans} utility and the Doppler shift was computed from 
the average wavelength shift of a few of the strongest lines in several orders. These
velocities are given in Table~\ref{tab:obs}.
 
\subsection{Stellar parameters and abundance calculations}
\label{sec:eqwidth}

\noi We derived chemical element abundances via a standard 
equivalent widths (EWs) analysis that closely follows  the procedures used in our 
previous works \citep[e.g.][]{Koch2009}. All analyses employed  the 2010 version of the 
stellar line analysis code MOOG~\citep{Sneden1973}. The EWs of the absorption 
lines were measured within the {\tt splot} task in IRAF assuming a Gaussian line profile. 
The line list was assembled from \cite{Moore1966} combined with the log $gf$ values from the Kurucz
database. The EWs for all the lines used in our analysis are given in Table~\ref{tab:lines}. In order to avoid saturation or spurious, weak lines, we conservatively excluded all Fe I and Fe II lines
 with EW$\geq$160$\rm m\AA$ and EW$\leq$40$\rm m\AA$ from our analysis. 

\begin{table}
\begin{center}
\begin{tabular}{lllrrrl}
\hline
& $\lambda$ & E.P.& & \multicolumn{3}{c}{EW [$m\AA$]} \\
\cline{5-7}
\raisebox{1.5ex}[-1.5ex]{X} & [$\AA$] & [eV] &\raisebox{1.5ex}[-1.5ex]{log\,$gf$} & UMI396 & UMI718 & UMI446 \\\hline                    
 O         &  6300.31       &   0.00   & $-$9.780    &  34.6      &   -       &     -         \\
 Mg        &  4571.10       &   0.00   & $-$5.623    &  -         &   123.6   &     93.5      \\
 Mg        &  4702.99       &   4.34   & $-$0.440    &  98.9      &   111.2   &     91.1      \\ 
 Mg        &  5528.41       &   4.34   & $-$0.498    &  132.4     &   107.8   &     -         \\    
 Mg        &  5711.09       &   4.34   & $-$1.724    &  36.3      &   -       &     -         \\
 Si        &  7423.50       &   5.62   & $-$0.314    &  30.8      &   -       &     -         \\
 Ca        &  4435.69       &   1.89   & $-$0.519    &  -         &   -       &     72.7      \\
 Ca        &  4454.78       &   1.90   &    0.260    &  -         &   116.0   &     -         \\ 
\end{tabular}           
\caption{ Equivalent width measurements.  Columns are: (I) Element; (2) Wavelength of the absorption line; (3) Excitation energy of the lower energy level; (4) Oscillator strength; (5,6,7) Equivalent width of the line measured for each star. This table is available in its entirety in the electronic version of the Journal.}
\label{tab:lines}
\end{center}
\end{table}

\noi Although we do not expect the effect of hyperfine splitting to be large in our spectra, they were investigated in the derivation of abundance ratios for Sc, Mn, Cu, and Ba, using wavelengths and log $gf$ values  taken
from the Kurucz database with the exceptions of the Sc II line at 467.0 from \cite{McWilliam1995} and 565.7, and 568.4 nm, calculated with the coefficients from \cite{Johnson2002}. Except for the larger corrections for  [Mn$/$Fe], 
which ranged from 0.09 to 0.15 dex, the inclusion of hyperfine splitting causes only small changes in the abundances of the other elements. Barium (scandium) were corrected by 0.05~(0.06)~dex in UMI718, 0.01~(0.03)~dex in UMI396 and 0.02~(0.02)~dex in UMI446. The copper abundance differed only in UMI718 (by 0.03~dex) with the inclusion of hyperfine splitting.

\noi Apart from a Non-LTE correction applied to the Al/Fe abundance by \cite{Cohen2010}, all of the data used in this paper are based on the assumption of LTE conditions. Our stars do not show any signs of ionisation imbalance in the abundances we calculate for Fe and Ti within the uncertainties. The calculations of ~\cite{Bergemann2008} for Cr and \cite{Bergemann2010} for Mn imply upward-corrections by $\sim$0.4 dex in these elements, but we note that their parameter range mainly included dwarf stars and did not extend to cool giants as in our sample. Thus we refrain from applying these corrections to these elements in the following.

\noi The initial estimates for the temperature and the surface gravity 
listed in Table~\ref{tab:obs} are the photometric values based on V-I, using the
calibration of \cite{Ramirez2005}, where we adopted a reddening of
E(B-V)=0.02 \citep{Schlegel1998} and an initial guess for the
metallicity of $\rm [Fe/H]=-2.0$. Throughout our analysis, we constructed model 
atmospheres for a large range of metallicities and microturbulent velocities by 
interpolation of  Kurucz's\footnote{\tt{http://kurucz.harvard.edu}} 
grid of one-dimensional 72-layer, plane-parallel, line-blanketed models  
without convective overshoot, assuming local thermodynamic equilibrium (LTE) 
for all species. For the UMi stars, we incorporated  the opacity distribution functions 
ODFNEW described by \citep{Castelli2003}\footnote{\tt\footnotesize{www.stsci.edu/hst/observatory/crds/castelli\_kurucz\_atlas.html}}.
As UMI446 was found to be very metal poor, we tested our final results with the $\alpha$ enhanced AODFNEW models as well but found that these did not affect the results more than 0.01~dex for any of the elements.

\noi These model atmospheres were then used in MOOG, together with the EWs to calculate 
the abundances of the  Fe I
and Fe II lines for each star. The iteration to obtain the best model parameters was made
first by checking the abundance trends with excitation potential in order to estimate the spectroscopic temperature and removing any trends with the reduced EW to estimate the microturbulence velocity. Finally, the best value for the surface gravity was measured by minimising the [Fe~I/Fe~II] ratio. During this procedure, all steps were iterated to convergence. Furthermore, we excluded strong lines with EW$\geq$160 to avoid saturated lines, and EW$\leq$40$\rm m\AA$ to avoid biasses due to low S/N (e.g., Magain 1984),  
  from this part of the analysis. Although it is difficult to calculate formal errors for the temperature, log~{\em{g}} and microturbulence as these parameters are highly interdependent (though see Section~\ref{sec:err}), our final models minimised the correlation coefficients for all of these (less than 6\% for T and V for all of our stars), and we could achieve a reasonable ionization balance solution, where the difference between the abundances of Fe I and Fe II are 0.07, 0.05, and 0.03 dex for UMI718, UMI396 and UMI446 respectively.

\noi In a few cases where the number of measured lines were very small for an element, we included lines with EW$<$40$\rm m\AA$ if the lines were clear and without blends. These exceptional, weaker lines ([Ce/Fe]  for UMI718, [Si/Fe], [Cr II/Fe II] and  [Ce/Fe] for UMI396, [Cu/Fe], [Y/Fe] and [Nd/Fe] for UMI446) were re-checked during the iterative process of finding the stellar atmosphere parameters which were normally adjusted through the trends seen for the [Fe$/$H] abundances. Despite the very low S/N in UMI446, there were features strong enough to be measured in both blue and red parts of the spectra.

\subsection{Chemical abundance measurements and uncertainties}
\label{sec:err}

\noi By varying the model atmosphere parameters by their 
typical uncertainties of 100K (T) and 0.2\,km/s (v$\rm _{mic}$), we find the systematic errors in
[Fe/H] to be 0.1~dex for UMI718, 0.15~dex for UMI718 and 0.13~dex for UMI446.

\noi The abundances calculated for 17 elements in our observed UMi stars are
presented in Table~\ref{tab:ab}, together with the Solar abundance scale, which 
was adopted from \cite{Asplund2009}. The detailed comparison with previous data
and our chemical evolution models are presented in Section~\ref{sec:model}.

\noi The overall abundance errors for each element given in
Table~\ref{tab:ab} are determined by using $\sigma/\sqrt{N}$ where $\sigma$
is the standard deviation of the individual line abundances and N is
the number of observed lines for this element. In Table~\ref{tab:ab}
we list each of these values.

\begin{table*}
\begin{center}
\begin{tabular}{l l r r c r r c r r r c}
\hline
 & SUN & \multicolumn{3}{|c|}{UMI718} & \multicolumn{3}{|c|} {UMI396} & \multicolumn{3}{|c|} {UMI446}  \\\hline
 X  &   log$\epsilon(X)_{\odot}$ &log$\epsilon(X)$   & $\sigma$(nr of lines) & [X$/$Fe] &log$\epsilon(X)$   & $\sigma$(nr of lines) & [X$/$Fe]&log$\epsilon(X)$   & $\sigma$(nr of lines) & [X$/$Fe]\\\hline
 Fe I & 7.5 & 5.38 & 0.17(88) & -2.12$\pm$0.02 & 5.37 & 0.19 (118) & -2.13$\pm$0.02 & 4.83&0.26(68) & -2.67$\pm$0.03  \\
 Fe II & 7.5 & 5.45 & 0.19(5) & -2.05$\pm$0.08 & 5.42 & 0.17(6) & -2.08$\pm$0.07 & 4.86 & 0.19(7)& -2.64$\pm$0.07 \\
 O I synt& 8.69 & - & - & - & 6.99  & $...(1)$ & $0.38\pm0.13$  & - & - & -\\
 Mg I   &  7.6  & $5.74$ & $0.09(3)$   & $0.26\pm0.05$   &  $5.79$  & $0.15(3)$ & $0.32\pm0.08$ & $5.26$ & $0.10(2)$   & $0.33\pm0.08$  \\
 Si I synt & 7.51 & - & - & - & $5.57$  & $...(1)$  & $0.19\pm0.13$  & - & - & - \\
 Ca I   &  6.34 & $4.55$  & $0.34(15)$   & $0.33\pm0.09$  &  $4.35$ & $0.15(16)$  & $0.14\pm0.04$  &  $3.97$ & $0.20(7)$  & $0.3\pm0.08$ \\
 Sc II$*$  &  3.15 & $1.14$  & $0.09(3)$    & $0.02\pm0.09$  &  $1.08$ & $0.19(4)$  & $-0.02\pm0.12$ &  $1.15$ & $0.27(3)$  & $0.63\pm0.18$ \\
 Ti I   &  4.95 & $2.96$  & $0.16(15)$   & $0.13\pm0.04 $ &  $2.86$ &  $0.19(25)$  & $0.04\pm0.04 $   &  $2.73$ & $0.23(7)$  & $0.45\pm0.09 $\\
 Ti II  &  4.95 & $3.07$  & $0.09(15)$   & $0.17\pm0.08$ &  $3.06$ &  $0.15(17)$  & $0.19\pm0.08$  &  $2.79$ & $0.34(13)$  & $0.48\pm0.12$ \\
 Cr I   &  5.64 & $3.50$  & $0.18(12)$   & $-0.02\pm0.05$  &  $3.27$  & $0.09(13)$  & $-0.24\pm0.03$ &  $2.99$ & $0.07(6)$   & $0.02\pm0.04$  \\
 Cr II$*$ & 5.64 & - & -& -& $3.33$ & $...(1)$   & $-0.23\pm0.13$  & - & - & - \ \\
 Mn I$*$   &  5.43 & $3.16$  & $0.17(5)$   & $-0.15\pm0.08$  &  $2.96$  & $0.12(3)$   & $-0.35\pm0.08$ &  $3.05$ &  $0.19(3)$   & $0.29\pm0.1$  \\
 Ni I   &  6.22 & $4.24$  & $0.17(17)$   & $0.14\pm0.04$ &  $4.11$ & $0.17(24)$  & $0.02\pm0.03$   &  $3.87$ & $0.11(5)$  & $0.32\pm0.05$\\
 Cu I$**$ & 4.19 & $1.92$  & $...(1)$   & $-0.18\pm0.1$  & - & - &  - & $1.64$ & $...(1)$   & $0.12\pm0.34$  \\
 Zn I   &  4.56 & $2.41$  & $0.12(2)$   & $-0.03\pm0.08$  & $2.54$ & $...(1)$   & $0.11\pm0.13$ & - & - & -  \\
 Y II   &  2.21 & $-0.32$ & $0.19(3)$  & $-0.48\pm0.13$  & $-0.35$ &  $0.04(3)$  & $-0.48\pm0.07$  & $-0.23$ & $...(1)$  & $0.18\pm0.34$ \\
 Ba II$*$  &  2.19 & $-0.46$ & $0.1(3)$  & $-0.55\pm0.1$ & $-0.35$ & $0.08(2)$  & $-0.45\pm0.09$ & $-1.21$ & $0.47(2)$  & $-0.74\pm0.34$   \\
 Ce II$*$& 1.58 & $-0.34$ & $0.03(2)$ & $0.13\pm0.1$ & $-0.65$ & $0.13(2)$ & $-0.15\pm0.12$  & - & - & - \ \\
 Nd II & 1.42 & - & -& -& $-0.68$  & $0.14(3)$ & $-0.02\pm0.1$ & $-0.31$  & $...(1)$  & $0.89\pm0.34$   \\
 Sm II$**$ &  0.96   & - & - & - & -0.27 & 0.04(3) & $0.9\pm0.07$ & - & - & -  \\\hline
\end{tabular}
\caption{Abundances of chemical elements in our target UMi stars and the Solar abundance scale, 
which was adopted from \protect\cite{Asplund2009}.($*$) For the values that come from the measurements of a single spectral line, we assume the abundance error to be the same as the largest calculated error for that star (0.1 for UMI718 and UMI396, 0.3 for UMI446). The abundance values are given in columns 5, 8 and 11 in terms of [X/Fe] and [Fe/H] for the Fe~I and Fe~II. ($**$)We measure a total Sm II abundance of [Sm/Fe]$=$0.88 for UMI396, with a surprisingly low standard deviation of 0.04~dex from the lines at 451.96, 452.39 and 456.62~nm. Although interesting, as all three lines are very weak with EW$\leq$25 and are at the blue end of the spectra, we omit this element in our interpretation of the chemical abundances in UMi.}
\label{tab:ab}
\end{center}
\end{table*}

\noi Similarly to the \cite{Kirby2012} data,  UMI446 and UMI718 have 
radial velocities which make the
telluric absorption lines coincide exactly with the forbidden oxygen
line at 630.03~nm. Therefore, this line was only measured in UMI396 by
fitting synthetic spectra.

\section{Abundances and the chemical evolution models for Ursa Minor}
\label{sec:model}

\noi In this section, we present the chemical abundances calculated for 
17 elements in three new UMi stars (see Table~\ref{tab:ab}) and use a set of homogeneous chemical evolution models to investigate 
the implications of the abundances in a data set comprising both our new data and those
obtained in previous studies of UMi. The main characteristics of our 
models are summarised in Table~\ref{tab:model_param}. 
The difference among the models is the treatment of the gas flows. Model A is a
closed box, without infall or outflow; model B has infall but no
galactic winds; in model C, we take into account both infall and
galactic winds. We constrain our models to follow the observational
SF given by \cite{Carrera2002}, while imposing a star
formation law given by

\begin{table}
\centering
\begin{tabular}[h]{|c|c|c|c|c|c|}
  \hline
Model & Description & Infall & Galactic winds & Final stellar &  \textbf{$M_{o}$} [$M_{\odot}$] \\
 & &  & & mass [$M_{\odot}$]& \\
\hline  
 A & Closed box & No  & No  & $\sim 2.3\times10^5$ & $1.2\times10^7$\\
 B & Accretion & Yes & No  & $\sim 2\times10^5$ & $2.4\times10^7$\\
 C & Acc + winds & Yes & Yes & $\sim 2\times10^5$ & $2.4\times10^7$\\
\end{tabular}
\caption{The parameters in the chemical evolution models. $M_{o}$ is the gas mass used in eq.~\protect{\ref{eq:infall}, and the laws for gas infall
and galactic winds are given in eq.~\protect{\ref{eq:infall}} and eq.~\protect{\ref{eq:wind}}, respectively.}}
\label{tab:model_param}
\end{table}

\begin{equation}
\psi(t)[10^{-4}\rm M_{\odot}/yr]= \left\{
  \begin{array}{l l}
    \begin{split}
    \Big[3\cdot10^{-2} M_{\rm gas}(t) \emph{N}^{\sigma,t_{0}}(t)\\
    - 10^{-10}t +0.5\Big]\\ 
    \end{split}
    &  \quad \rm if ~t \le 5\rm Gyr\\
    \begin{split}
       0 \\
       \end{split}
& \quad \rm if ~t> 5\rm Gyr
  \end{array} \right\}
\label{eq:sfr}
\end{equation}

\noi where, \emph{N}$^{\sigma,t_{0}}$ is a normalized Gaussian
function with $\sigma$ of 0.5~Gyr, centered at $t_{0}$ equal to 0.5~Gyr. The SF law
provided by Eq.~\ref{eq:sfr} mimics the observed peak of early SF with a subsequent period of extended SF with
lower efficiency, which stops after 5~Gyr as proposed by
\citet{Carrera2002}.

\noi The infall law used in models B and C takes the same form as the 
peak of the SF described in Eq.\ref{eq:sfr} and is given by
\begin{equation}
G_{\rm Infall}(t)[M_\odot/yr]=M_{o}\cdot \emph{N}^{\sigma,t_{0}}(t)
\label{eq:infall}
\end{equation}
Models B and C have no initial gas and reach $M_{o}$ which is given in Table~\ref{tab:model_param} following the infall law given in Eq.~\ref{eq:infall}. Although the infall law is not directly 
included in the SF law, the common form of these equations 
provides an implicit  connection between the SF and the infall.

\noi The galactic wind for model C has the simple form 
\begin{equation}
  G_{\rm wind}(t) [10^{-4}M_{\odot}/yr]= \left\{
  \begin{array}{l l}
    0 & \quad \rm if ~t \le 2Gyr\\
    2 \cdot 10^{-5}\cdot M_{gas}(t) & \quad \rm if ~t > 2Gyr
  \end{array} \right.
\label{eq:wind}
\end{equation}
and is assumed to act only at the end of the gas infall and of the main SF episode (after 2~Gyrs).
By implementing the winds with the simplest possible assumptions,
we are able to investigate their implication on the chemical evolution,
without being forced to constrain their origin(s), given the fact their origins
 are  still not clearly understood.

\begin{figure*}
  \includegraphics[scale=0.51]{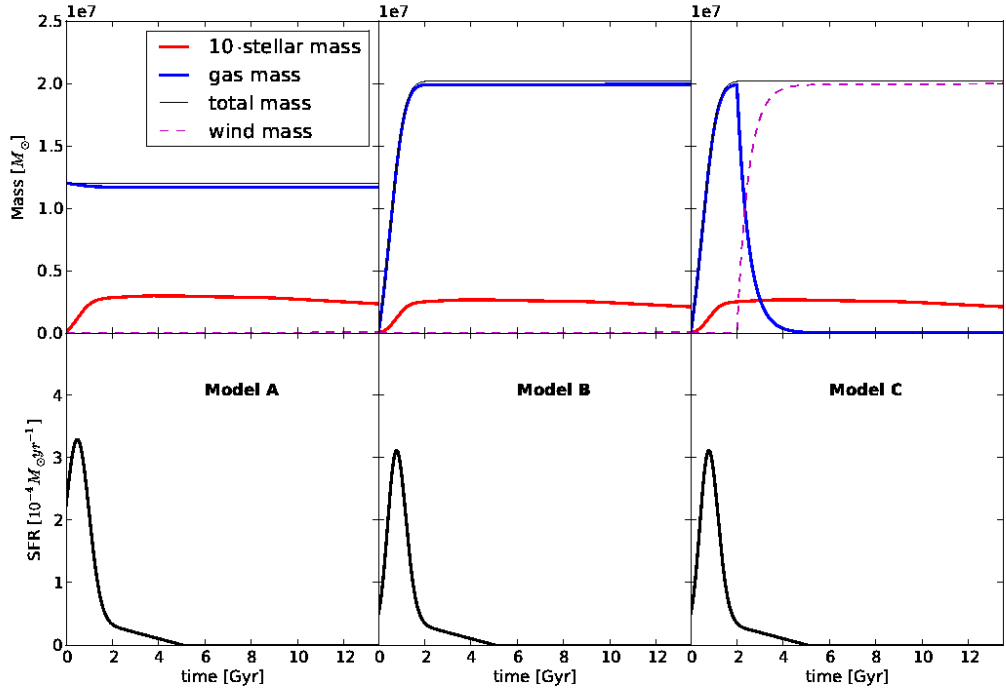}
\caption{ Top: The mass of baryons, gas, stars and
wind as functions of time for models A, B and C (from left to right). 
Bottom: the SF rate as a function of time for the same models.}
\label{fig:gas}
\end{figure*}  

\noi The top row in Fig.~\ref{fig:gas} shows the total mass, gas mass and
the stellar mass for the three models as a function of time, while in the second row
  we show the three SF histories obtained for each individual model. 
As model A, which has all the gas available from the outset, produces stars 
more efficiently than models B and C, in order to obtain the same final stellar masses 
in all models, the initial gas mass in models B and C is twice that of model A.
Both models A and B have a large amount of gas left at the end, however, and it is clear that
the loss of gas through winds is necessary to explain the lack of gas observed in UMi.

\noi In all of our models the IMF of \cite{Salpeter1955} is used, and
the other chemical evolution parameters, such as the treatment of SNe~Ia, 
and the nucleosynthesis, as well as the stellar lifetimes are those adopted by~\cite{Francois2004}. 
The exceptions are Mn, for which we follow the nucleosynthesis
adopted in \cite{Cescutti2008}, and the neutron capture elements
Ba and Y, for which we adopt the same nucleosynthesis of model EC$+$s 2 shown in \cite{Cescutti2014}.
The latter study assumed that the r-process elements are produced by electron capture SN in stars with masses
between  $\rm 8-10 M_\odot$\, and the s-process elements are produced in rotating massive stars (spinstars)
with masses between $\rm 15\, and\, 40 M_\odot$.

\begin{figure}
  \includegraphics[scale=0.36]{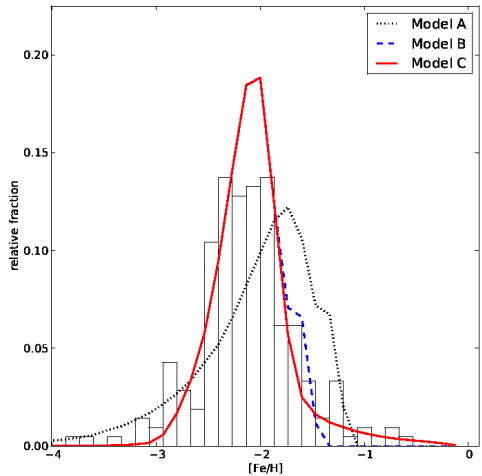}
\caption{Comparison between the observed MDF histogram~\citep{Kirby2011}
and the theoretical MDF produced by our models A (black), B(blue) and C(red).}
\label{fig:MDF}
\end{figure}  

\noi The parameters we assumed aim to reproduce the
observed metallicity distribution function (MDF) observed by \cite{Kirby2011}.  
Fig.~\ref{fig:MDF} compares the MDF found in our models with the observed distribution of stellar
metallicities. By means of a simplified chemical evolution model, \citet{Kirby2011} already showed that
it is not possible to find a model which is able to reproduce the observed MDF
 without infall. Our findings agree with their conclusions, and indeed 
our model A only poorly approximates the observed MDF.  
While our model B (which includes gas infall) is able to match the main characteristics 
of the observed MDF, as shown by Fig.~\ref{fig:gas}, it  
cannot explain the absence of gas displayed at the present time by UMi -
 which is also a common feature of all the dSph galaxies around the Milky Way.
 The implementation of a galactic wind in Model C, albeit in a simplified manner, solves this problem. We note 
that it also improves the match of the theoretical MDF
with the observed one by slightly extending the tail of the distribution
at high metallicity. We emphasize that the MDF and the SF law are the only observational constraints that were used in setting the constants in Eqs.~\ref{eq:sfr},\ref{eq:infall} and \ref{eq:wind}. It is then noteworthy that the ensueing chemical evolution model, shown in Figs.~\ref{fig:alpha}-\ref{fig:ybamodel} are successful in matching the observed abundance ratios, even though they were not used in fitting the model parameters.

\subsection{$\alpha$-elements: O, Mg, Si, Ti, Ca}

\begin{figure*}
\begin{minipage}[]{0.99\linewidth}
  \includegraphics[scale=0.45]{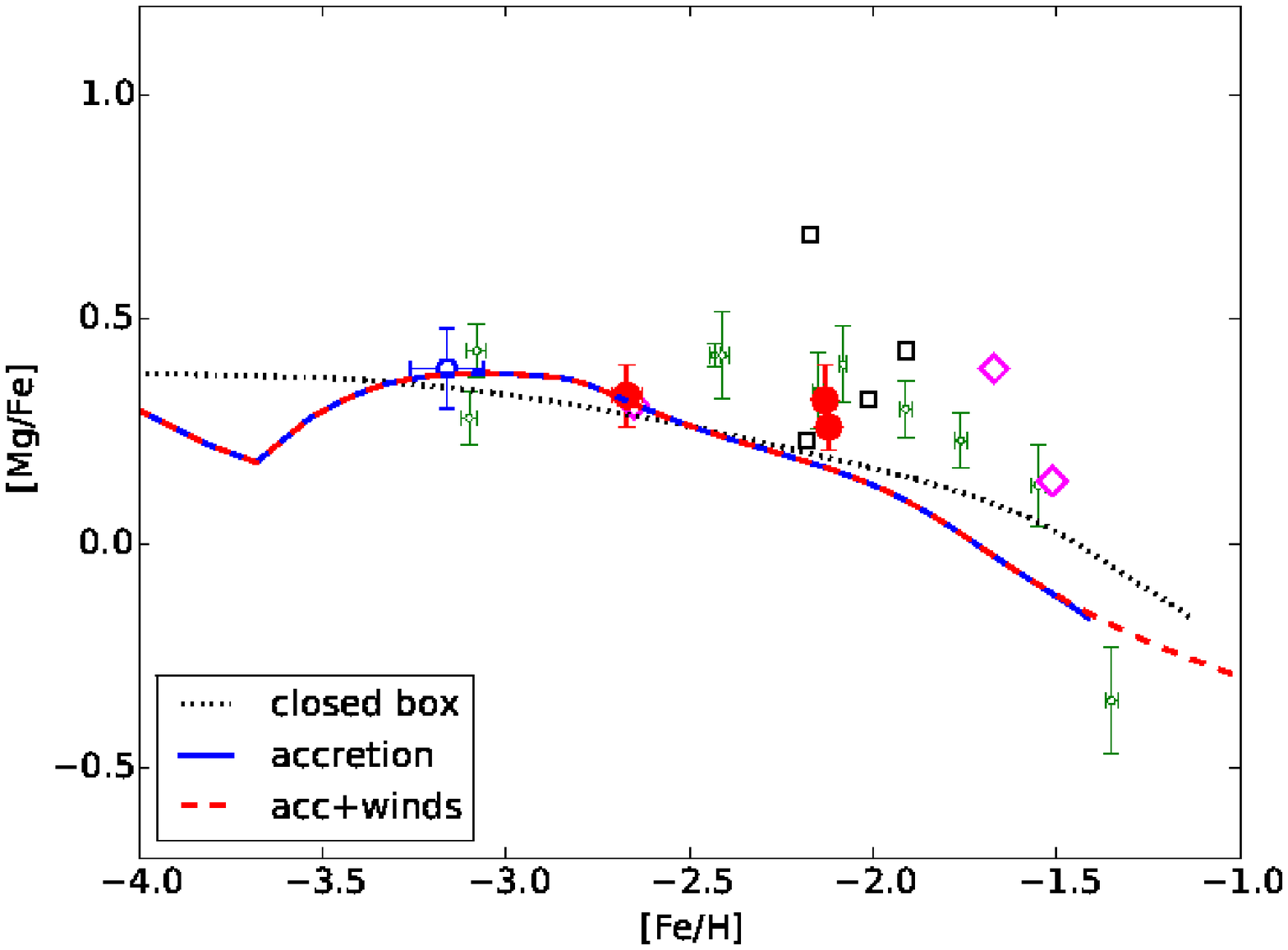}
  \includegraphics[scale=0.45]{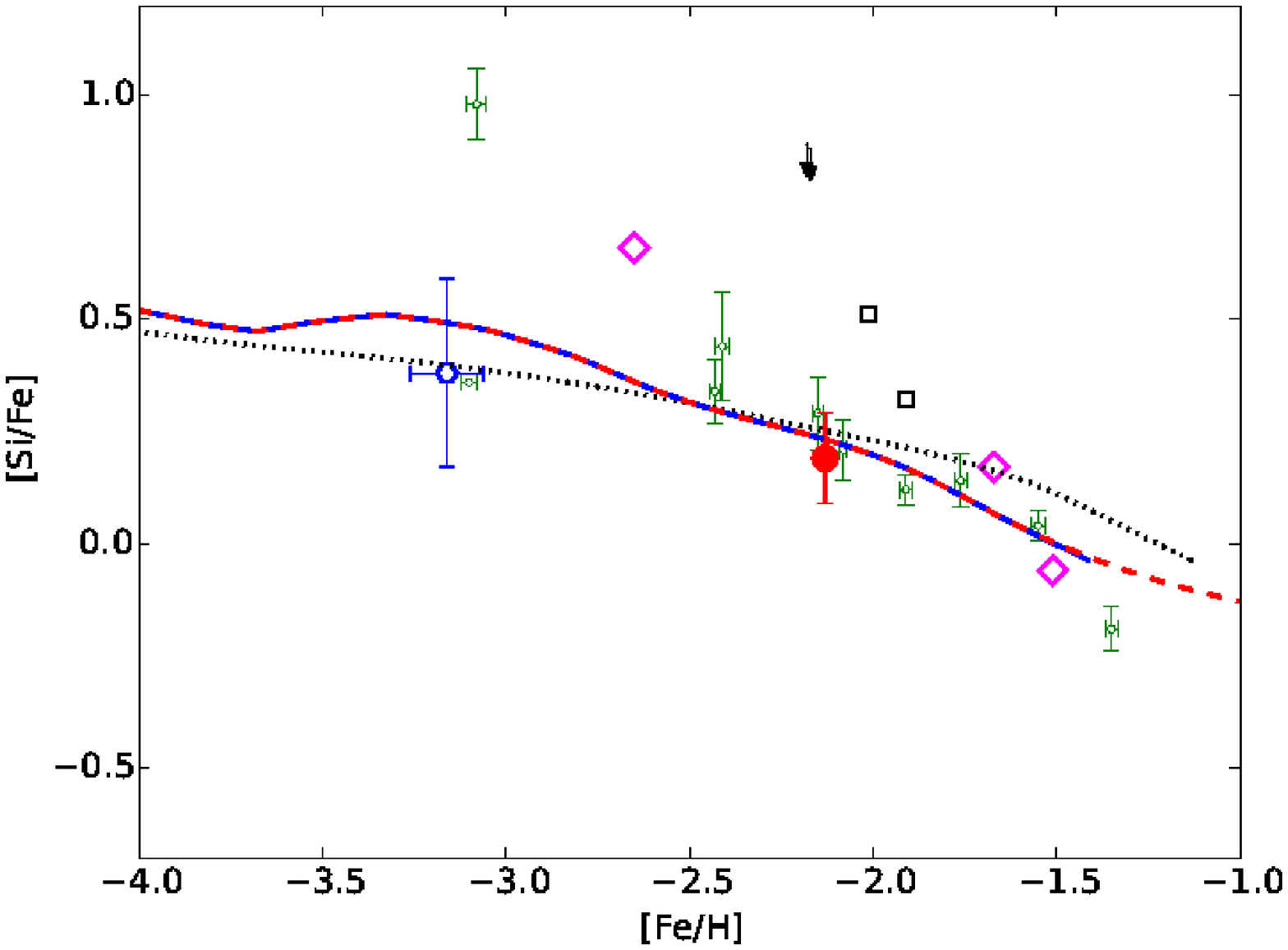}
\end{minipage}
\begin{minipage}[]{0.99\linewidth}
  \includegraphics[scale=0.45]{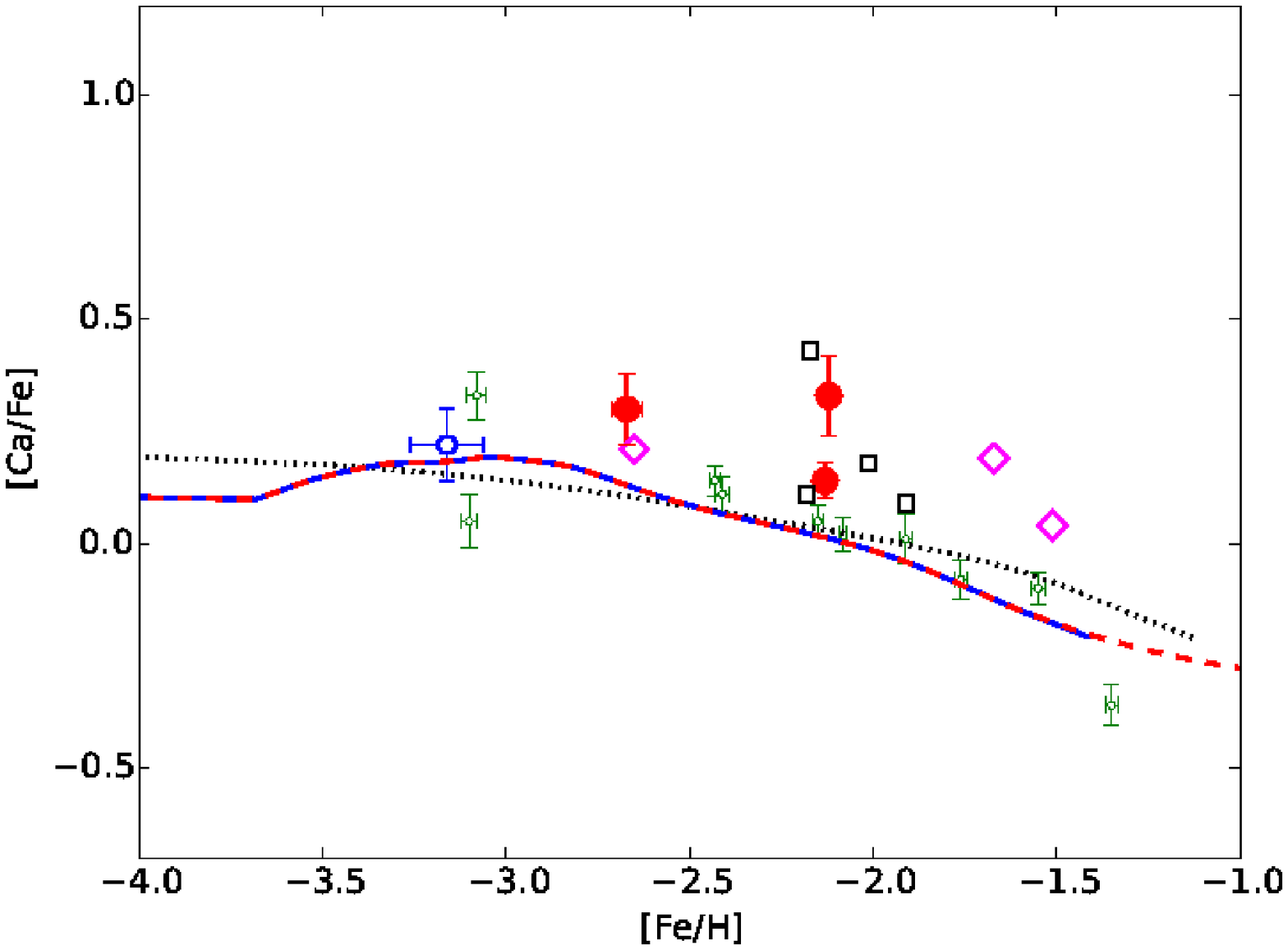}
  \includegraphics[scale=0.45]{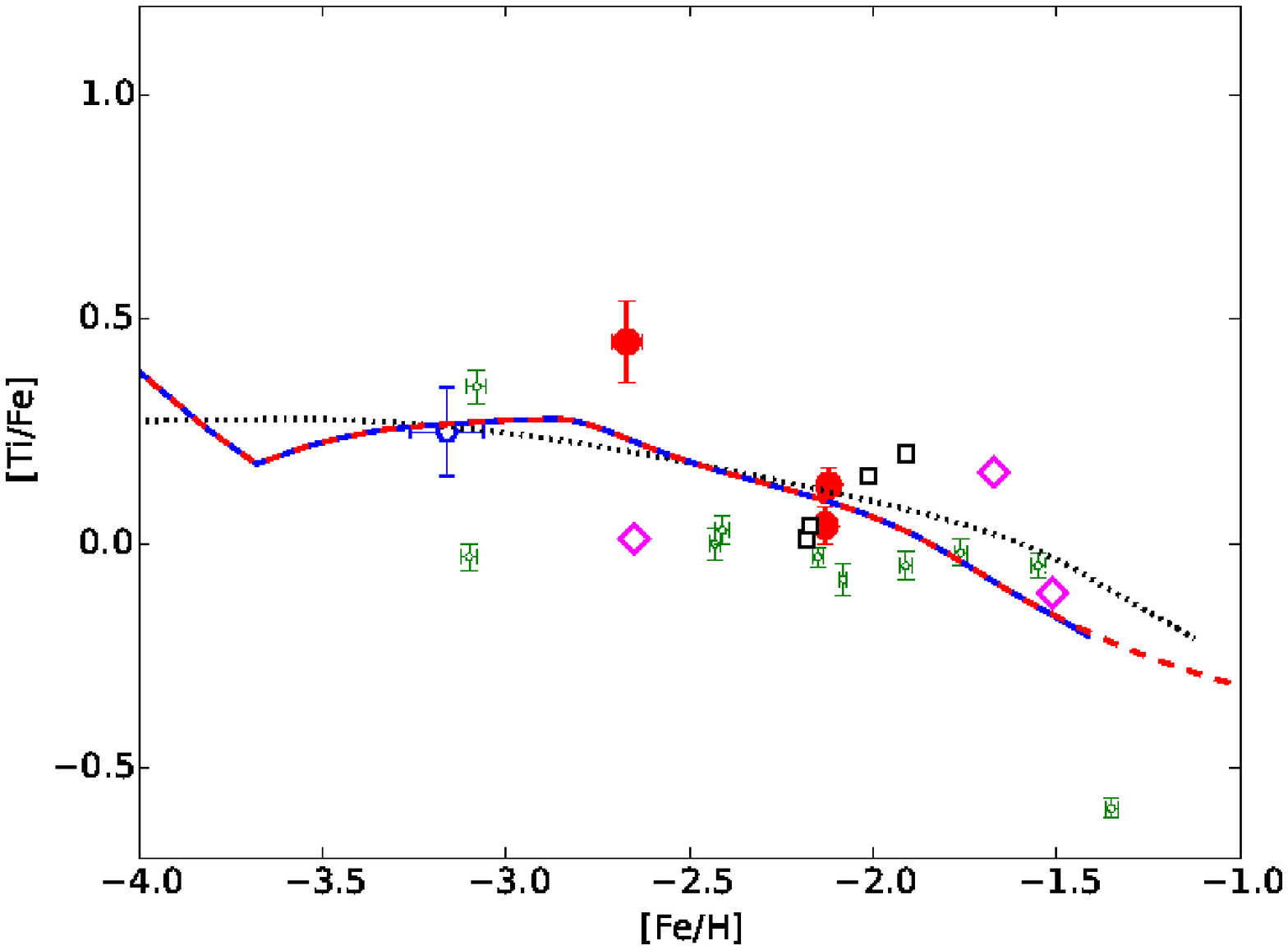}
\end{minipage}
\caption{$\alpha$-elements in UMi. The symbols are data from: this work (red filled circles), \protect\cite{Shetrone2001} (black squares, and arrows for the upper limits) and \protect\cite{Cohen2010} (green dots), \protect\cite{Sadakane2004} (empty magenta diamonds) and \protect\cite{Kirby2012} (blue empty hexagon). The average abundances for these elements in halo stars as calculated by \protect\cite{Cayrel2004} are: [Mg/Fe]=0.26~dex, [Si/Fe]=~0.42~dex, [Ca/Fe]=0.35~dex, [Ti/Fe]=0.26~dex.}
\label{fig:alpha}
\end{figure*}
 
\begin{figure}
\begin{center}
  \includegraphics[scale=0.44]{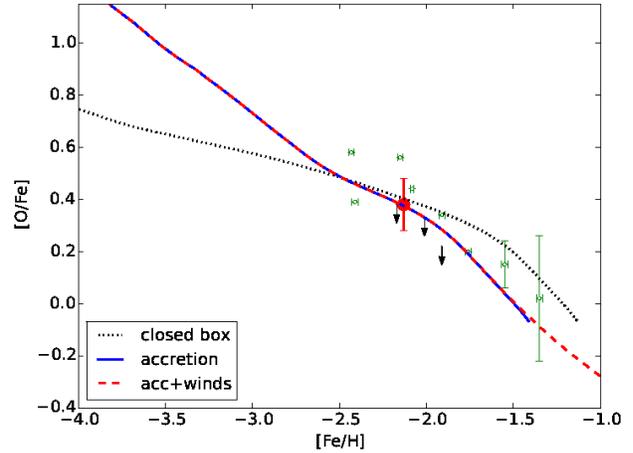}
\end{center}
\caption{Oxygen abundances in 12 UMi stars . Literature data are taken from \protect\cite{Shetrone2001} (black arrows for the upper limits) and
 \protect\cite{Cohen2010} (green dots). Our star is shown with the red filled circle. The chemical evolution models are the closed box (black dotted line), model with 5~Gyr of gas infall (blue solid line) and the model with infall and gas loss through winds (red dashed line). }
\label{fig:o}
\end{figure}
 \noi The measured $\alpha$-element abundances (see
 Fig.~\ref{fig:alpha}) are generally consistent with a  SF efficiency that is 
lower than that of the Galactic halo, as also shown by our models which show a 
gradual decrease in the [$\alpha/$Fe] caused by the onset of SNIa already at very low metallicities 
(in the models at [Fe/H]$=-3$). On the other hand, the data for both [Mg/Fe] 
and [Ca/Fe] indicate a plateau at [Fe/H]$\leq-2$~dex though sparsely sampled, which 
is not found in the models. For the rest of the individual elements, our [Ca/Fe] and [Ti/Fe] 
abundances, which are closer to \cite{Shetrone2001} and \cite{Sadakane2004} data at higher 
[X/Fe] values than the \cite{Cohen2010} data, are only slightly lower than the halo abundances. The 
[Si/Fe] ratio has a larger scatter in UMi than in the halo (where it is nearly constant), and 
the [O/Fe] ratio which is reproduced well by our models (see Fig.~\ref{fig:o}) is lower in UMi 
than in the halo stars~\citep{Cayrel2004,Cohen2010}. However, additional lower metallicity stars are needed
in order to better distinguish among the different models.

\noi As seen in Fig.~\ref{fig:alpha}, model A (without infall and winds), tends to
  have a less pronounced decrease at higher [Fe/H] which is due 
to the relatively higher SF compared to models B and C. Concerning these two models, the only difference is that 
  model C (with winds) is able to produce stars at a higher
  metallicity. This  reflects the MDF  we obtain for it, driven
  by the late metal pollution  (mostly  SN~Ia) in a decreasing reservoir
  of gas which produces a higher concentration of metals.  On the other
  hand, we underline that while the winds do not change significantly the
  chemical evolution for this galaxy, they help explain the absence of
  gas at the present time and improve the match with the MDF.  
  Instead of evoking an episodic star formation, our models use an extended, low-efficiency star formation period.
  
\subsection{Fe-peak elements: Sc, Cr, Ni, Mn, Cu, Zn}
\label{sec:Fe} 

\begin{figure*}
    \begin{minipage}{0.48\linewidth}
   \includegraphics[scale=0.45]{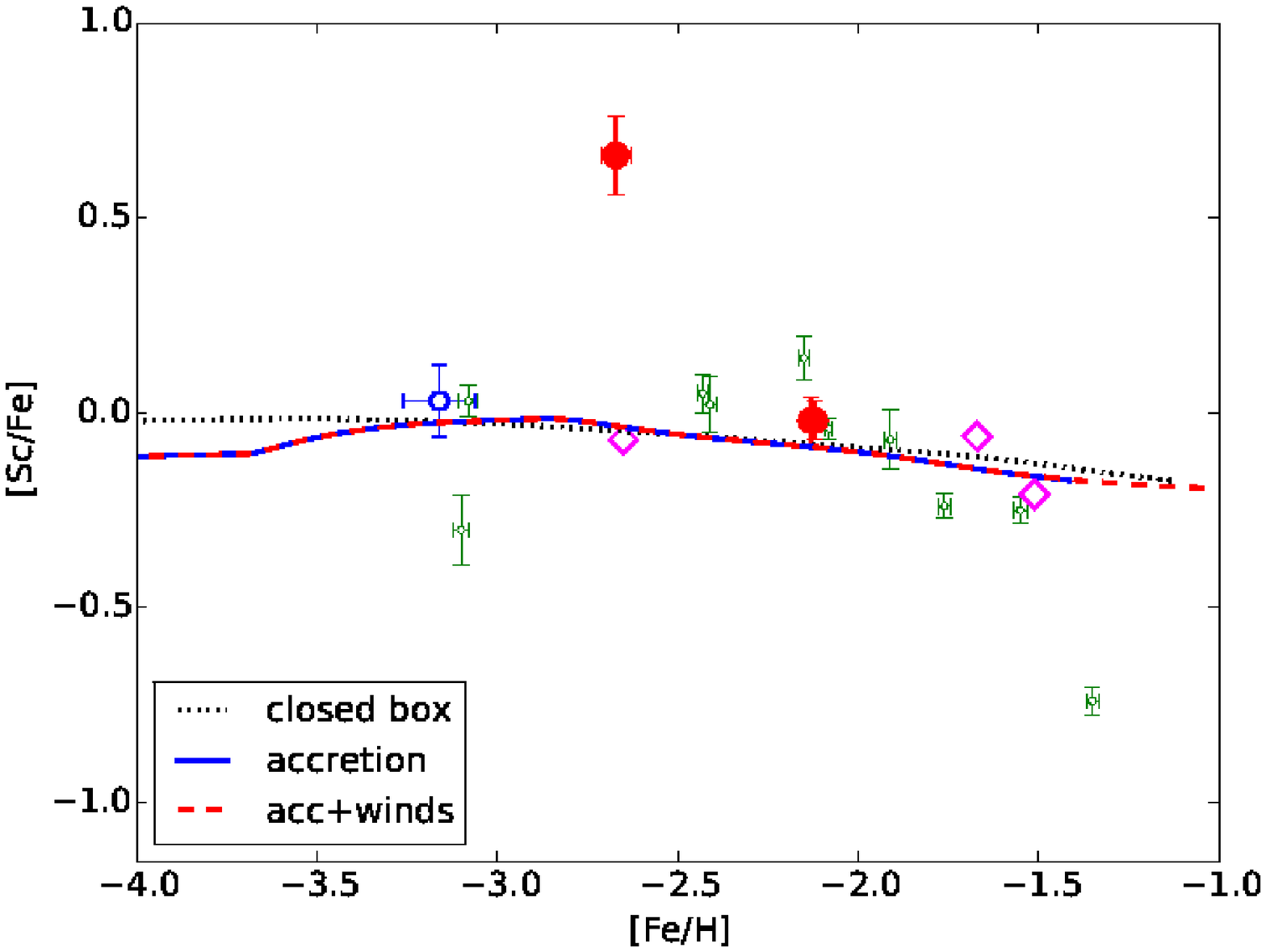}
\end{minipage}
 \hspace{0.5cm}
   \begin{minipage}[h]{0.48\linewidth}
   \includegraphics[scale=0.45]{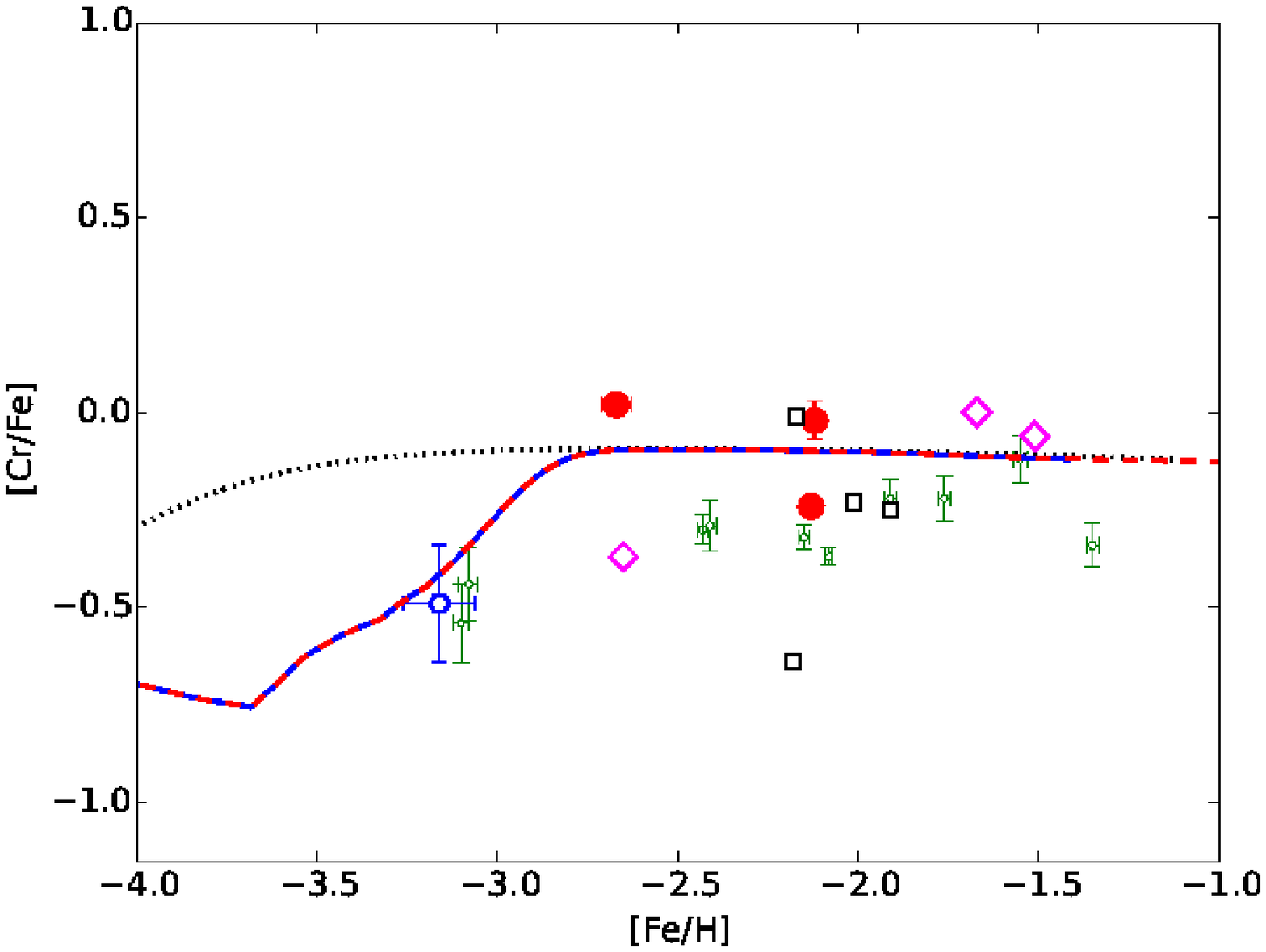}
\end{minipage}
 \hspace{0.5cm}
    \begin{minipage}[h]{0.48\linewidth}
   \includegraphics[scale=0.45]{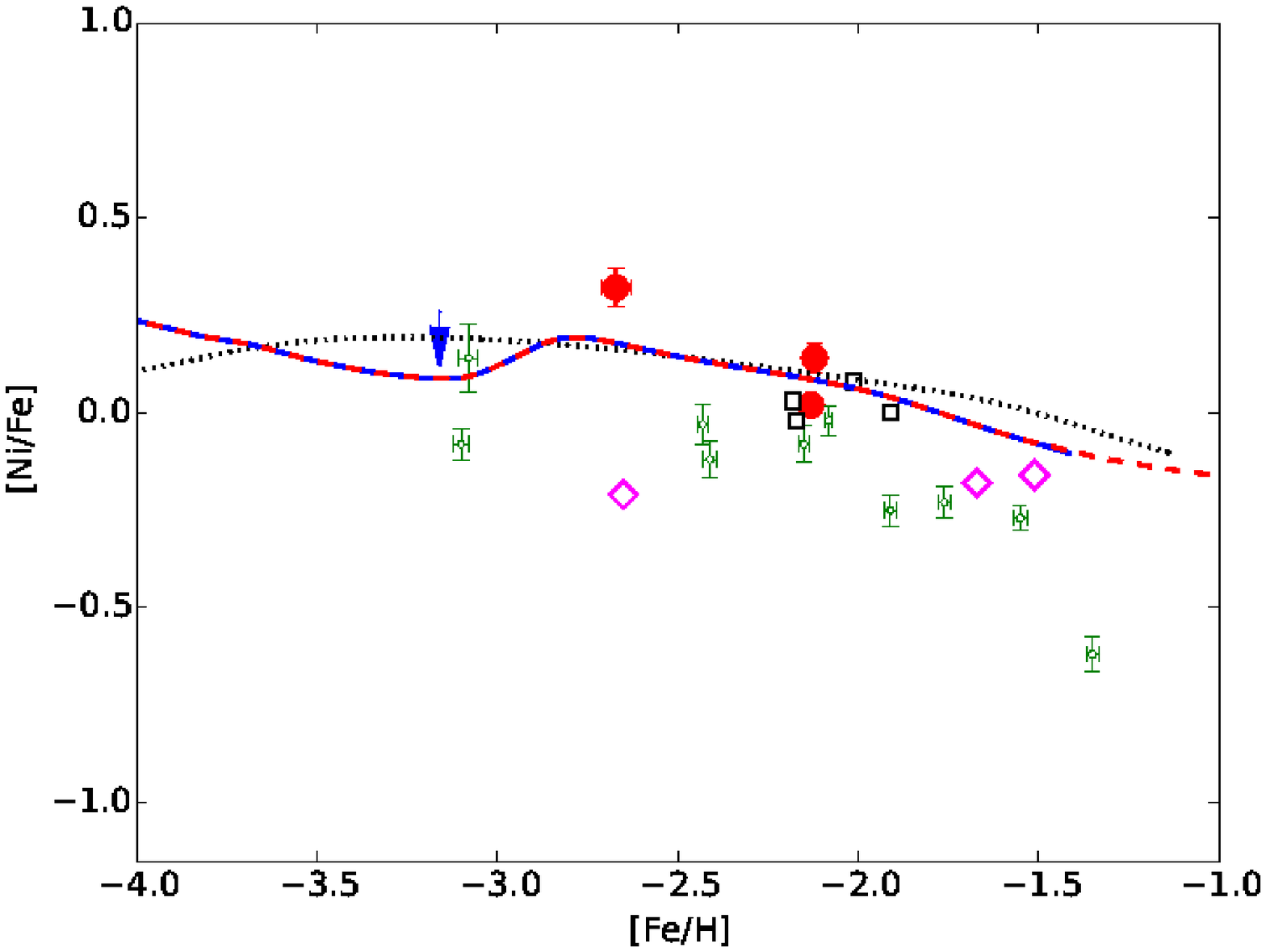}
\end{minipage}
 \hspace{0.5cm}
\begin{minipage}[hl]{0.48\linewidth}
    \includegraphics[scale=0.45]{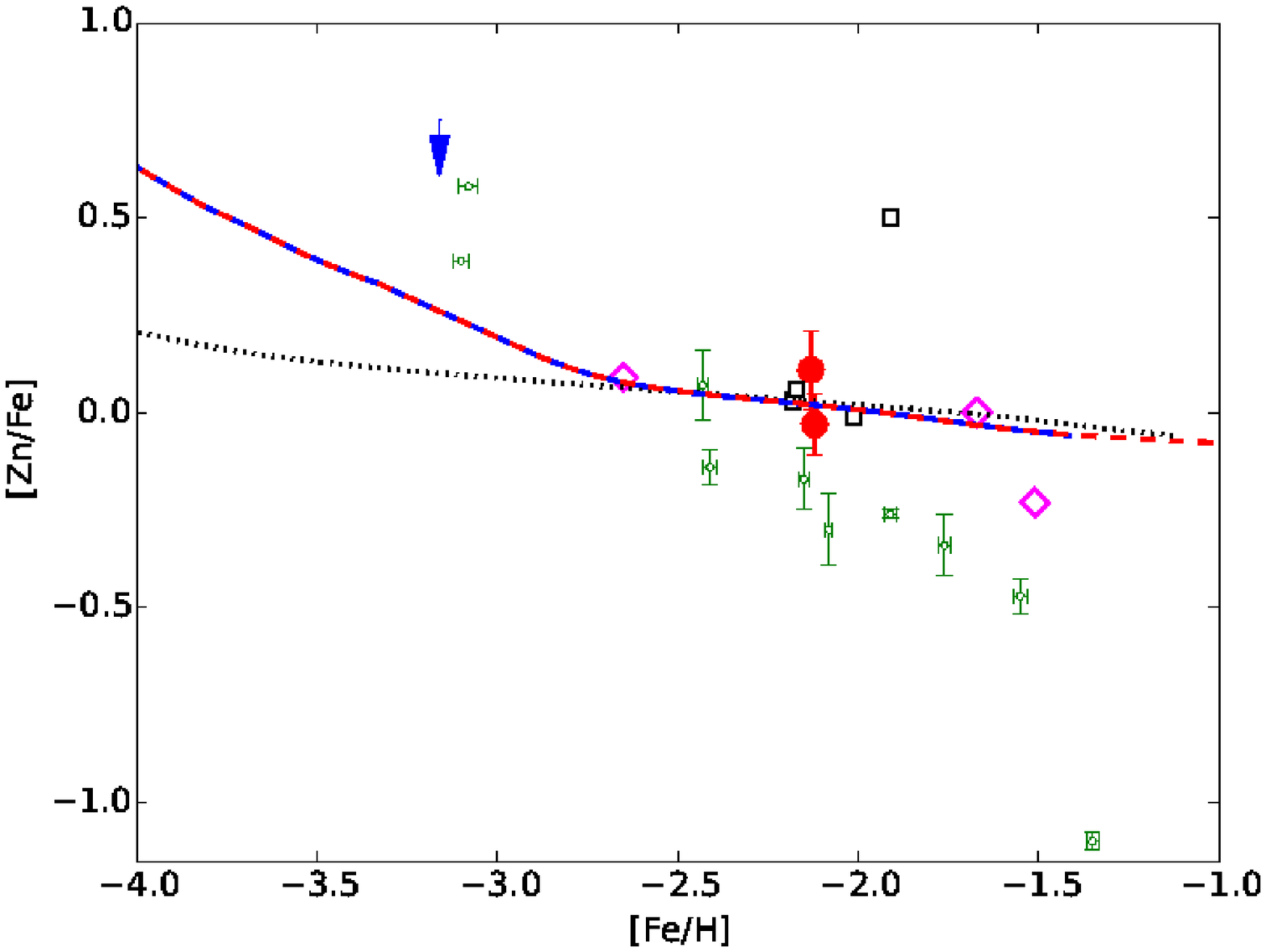}
\end{minipage}
 \hspace{0.5cm}
   \begin{minipage}[h]{0.48\linewidth}
    \includegraphics[scale=0.45]{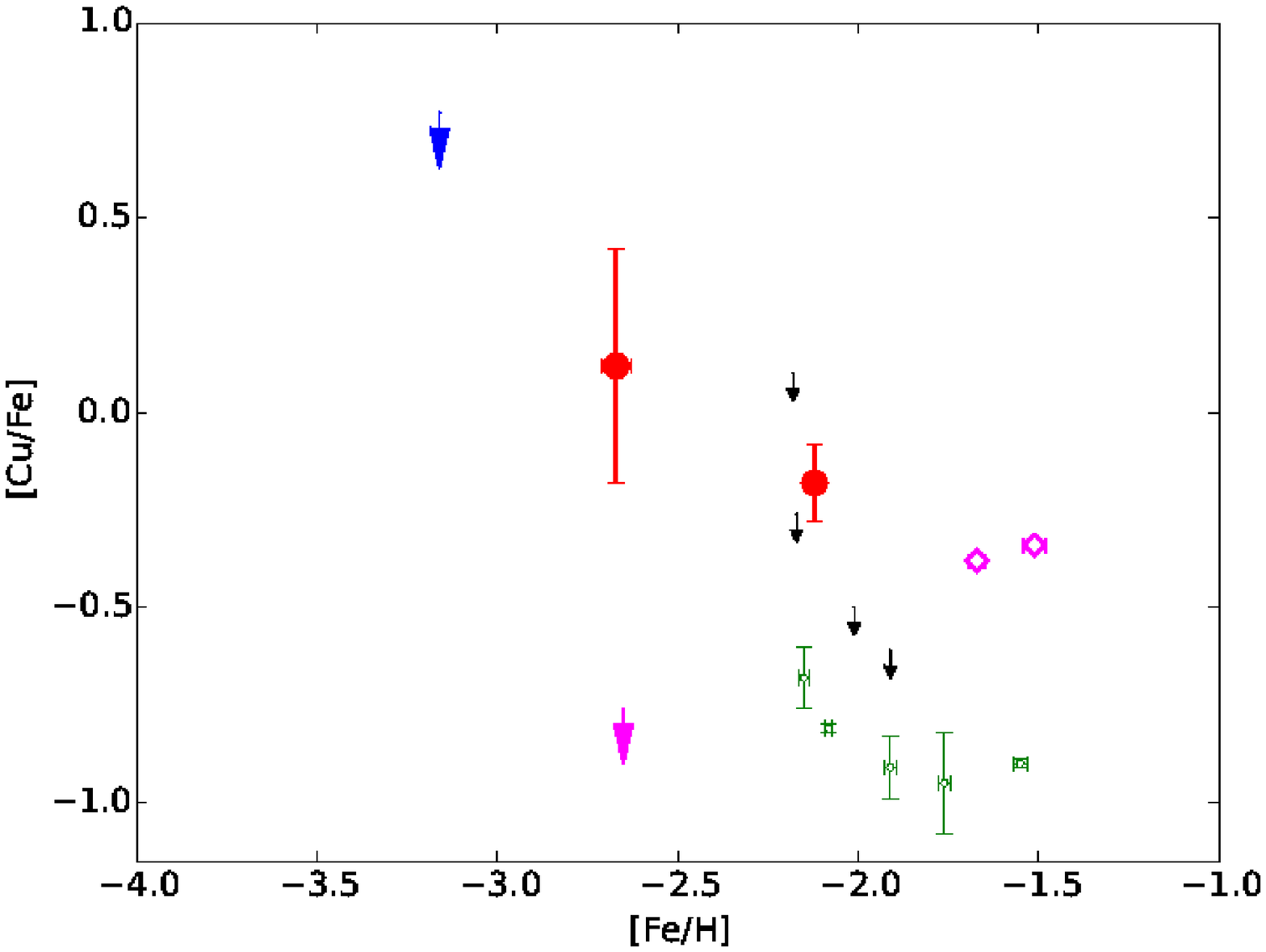}
\end{minipage}
 \hspace{0.5cm}
   \begin{minipage}[h]{0.48\linewidth}
    \includegraphics[scale=0.45]{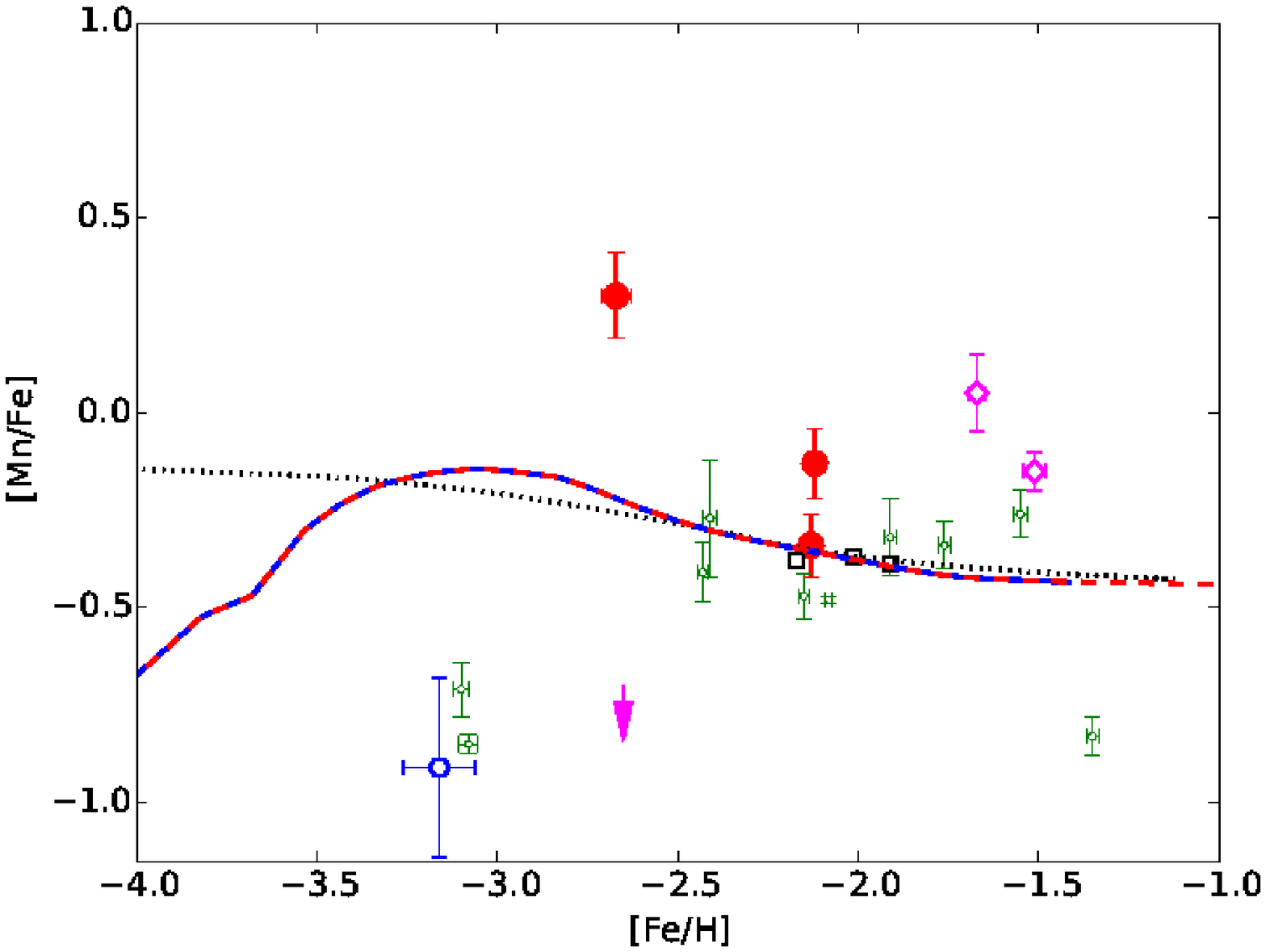}
\end{minipage}
\caption{Abundances for the Fe-peak elements. The symbols and colors are the same as in  Fig.~\ref{fig:alpha}.}
\label{fig:ironpeak}
\end{figure*}  

\noi As seen in Fig.~\ref{fig:ironpeak}, our models generally reproduce the abundances of the Fe-peak elements well. Similarly to the \cite{Sadakane2004} data, our stars have abundances close to those found in halo giants~\citep{Cayrel2004} in Ni and Cr (slightly higher in Zn), and are higher compared to those from \cite{Cohen2010}. The [Cr/Fe] abundance at nearly the solar value in UMI446, UMI718, UMI K from \cite{Shetrone2001} as well as the two more metal rich \cite{Sadakane2004} stars shows a plateau for this element apart from the decline towards lower metallicities which is likely to be caused by NLTE conditions as argued by \cite{Bergemann2010}. The Cu abundances we measure in UMI718 and UMI396 are in agreement with the whole UMi sample available to date, which show a decline with increasing metallicity.

\noi  For Mn, we match the bulk of the data at intermediate metallicity, so the model 
is relatively successful, however the decrease predicted at the extremely low metallicities
by our models is not consistent with the data in which this signature is detected at [Fe/H]=-3~dex. 
Our most metal poor star, UMI446, deviates from rest of the data set with a strong enhancement 
in Fe-peak elements, particularly in Sc and Mn. Although stars measured
by \cite{Sadakane2004} display [Mn/Fe]$>$0 as well, UMI446 has a much lower [Fe/H] than these two,  
and cannot be explained by our chemical evolution model even if we take into account metal-dependent Mn yields for both SNe Ia and SN II as used by \cite{Cescutti2008} and \cite{North2012}. 

\noi As a final remark in this subsection, we find the chemistry 
shown by the most metal rich star measured by \cite{Cohen2010} challenging.
An anomalous pollution by SN~Ia which cannot be simulated by our simple 
chemical evolution models, could explain the measured
 drop in the ratio of $\alpha$-elements (and neutron capture elements). 
However, it is more difficult to explain the strong enrichment of Fe 
in this star without an enrichment of the other Fe-peak elements.

\subsection{Neutron capture elements:  Y, Ba, Ce, Nd}

\begin{figure*}
    \begin{minipage}[h]{0.48\linewidth}
        \centering
  \includegraphics[scale=0.45]{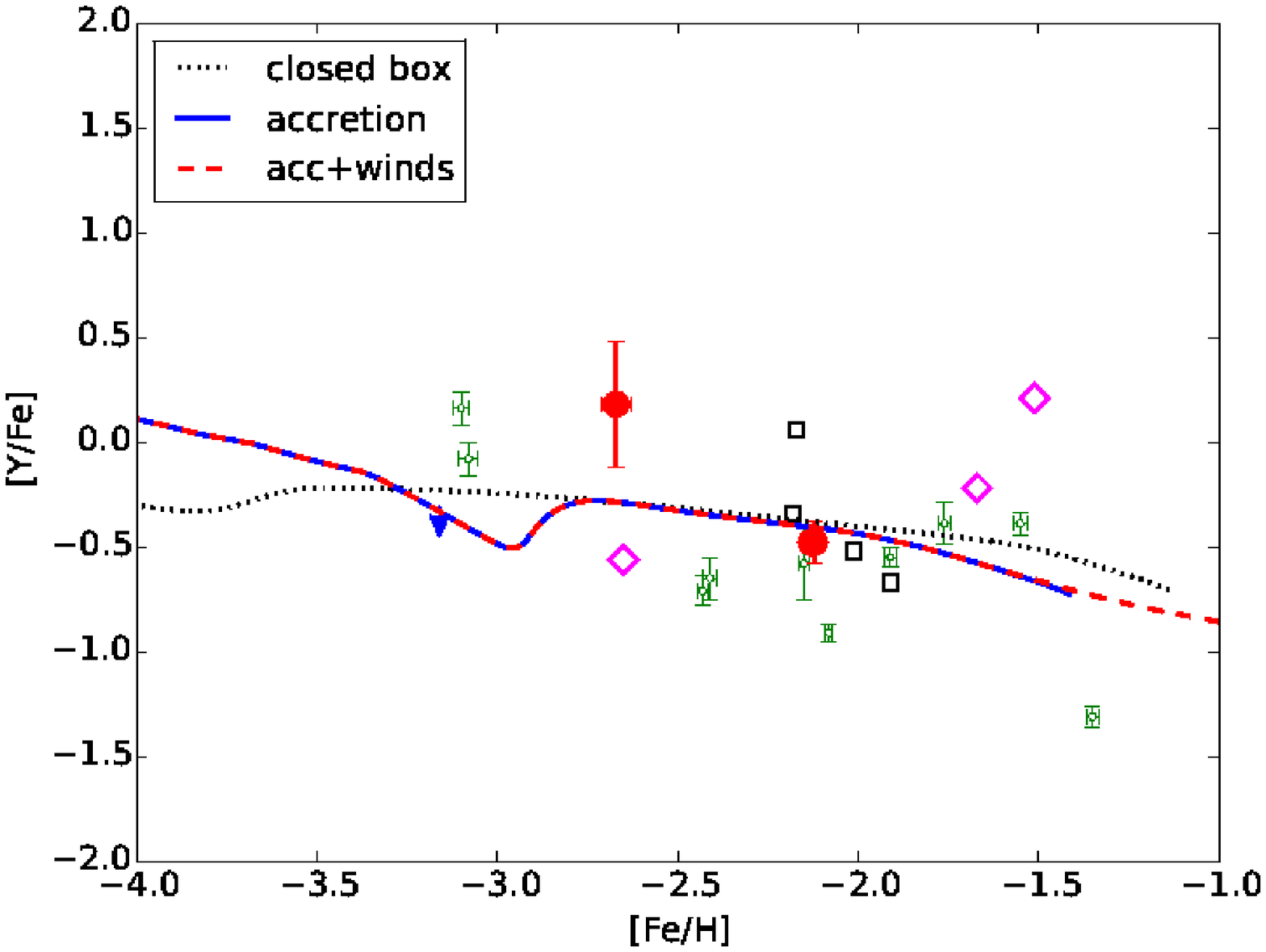}
\end{minipage}
 \hspace{0.5cm}
\begin{minipage}[hl]{0.48\linewidth}
  \centering 
  \includegraphics[scale=0.45]{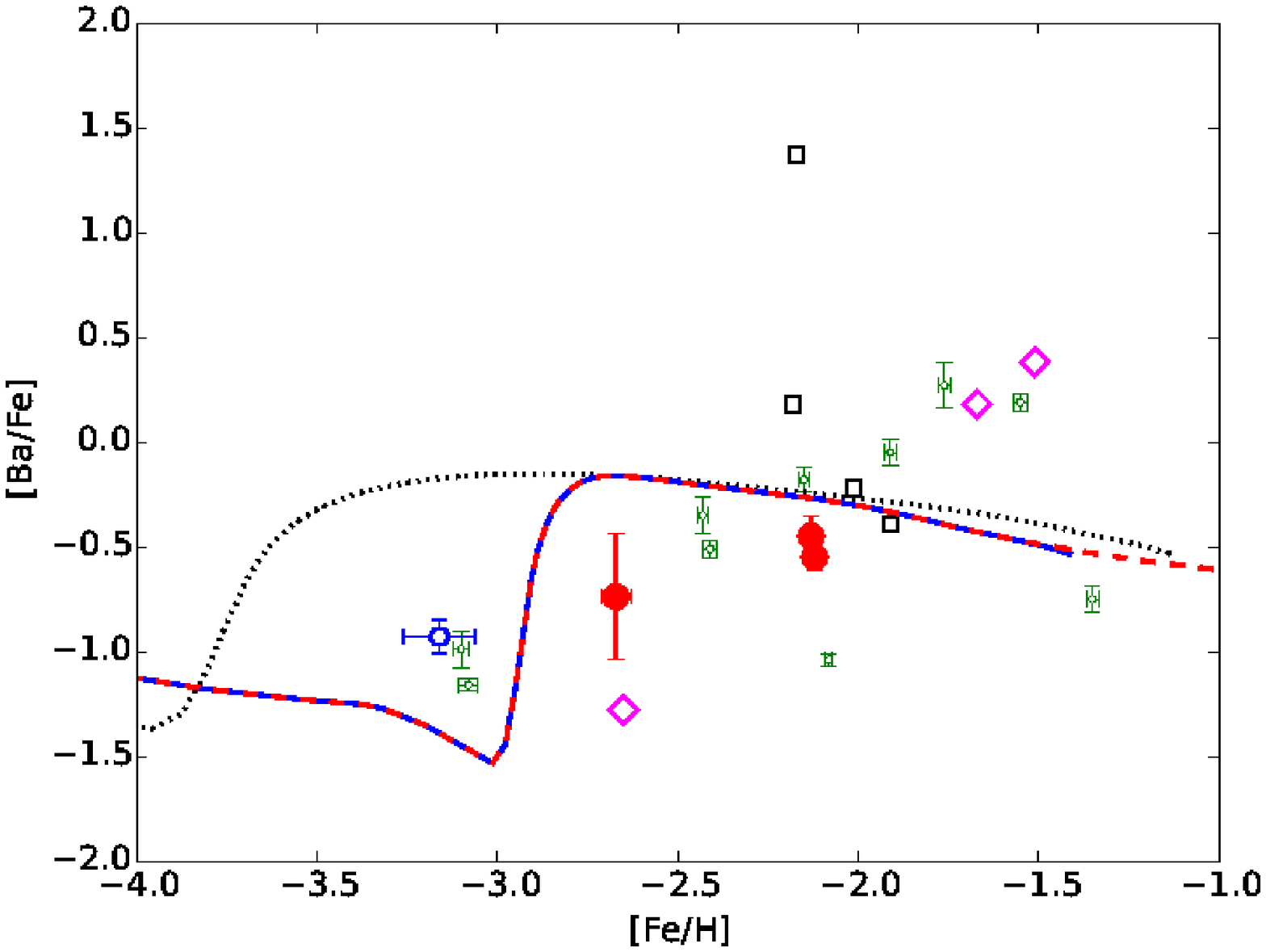}
\end{minipage}
\caption{Abundances for the neutron capture elements in UMi giants.The symbols and colors are the same as in Fig.~\ref{fig:alpha}}
\label{fig:neutron}
\end{figure*}

\begin{figure}
\begin{center}
 \includegraphics[width=0.5\textwidth]{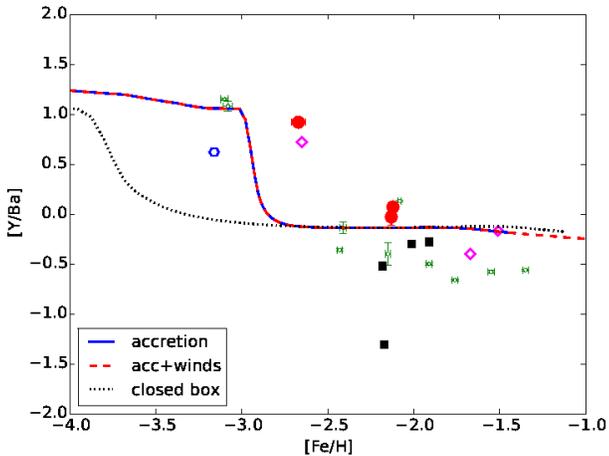}
 \caption{The ratio of yttrium to barium abundances in the UMi stars. The symbols and the colors are as in  Fig.~\ref{fig:alpha}.}
\label{fig:ybamodel}
\end{center}
\end{figure}  

\noi There are several interesting points that are seen in the neutron capture element abundances in UMi. 
First, all of the neutron capture elements that we measure have very large scatter across the 
observed stars, covering a range of about 2~dex. The Y and Ba abundances for the data set are shown in
Fig.~\ref{fig:neutron} and the [Ce/Fe] and [Nd/Fe] abundances in
two new stars each are given in Table~\ref{tab:ab}. Due to this spread in the observed abundances, 
we expect that inhomogeneous modeling would
be required to obtain a more precise match to the data (see for example Cescutti 2008,
Cescutti et al. 2013 and Cescutti and Chiappini 2014).
Nevertheless, a simple chemical evolution model should be able to reproduce the mean
trend. 

\noi The second point of note is the increase of [Ba/Fe] with metallicity seen in 
the entire UMi sample (Fig.~\ref{fig:neutron}) and which overlaps 
with the halo stars~\citep{Fulbright2000,Cohen2010}. In Fig. \ref{fig:neutron}, we show that
the model results for Y and Ba are in reasonable agreement with the data in that there is a systematically lower average [Ba/Fe] ratio at [Fe/H]$<\leq$3~dex compared to higher metallicities. In addition, although the present data in UMi cannot resolve these details, the models predict a dip in both
the [Y/Fe] and [Ba/Fe] abundance ratios at
[Fe/H]$ \sim-3$, which is due to the timescale at which the r-process
production by electron capture SN (see \cite{Cescutti2013} for
details) starts. 

\noi Finally, as predicted by the theoretical nucleosynthesis prescriptions we use, the data show
a much lower [Ba/Fe] ratio at very low metallicity, where the [Y/Fe] remains approximately at the same level.
At lower metallicity, the production of these neutron capture elements could be sustained by s-process production in fast rotating stars. Note that the low [Ba/Fe] abundance ratio is also found in the Hercules dSph over a broad range in [Fe/H]~\citep{Koch2013}
This effect is clearer when we plot the ratio between these two neutron capture elements
 (see Fig.~\ref{fig:ybamodel}). In this model, before the r-process production starts at [Fe/H]$\sim-$3,
the ISM is enriched only through s-process production in spinstars, which is confirmed by an overabundance in [Y/Ba]
found in the data at very low metallicities. There is possibly an offset of $\sim$0.5~dex
between the last data with high [Y/Ba] defined by UMI446 from this work and COS4 from~\cite{Sadakane2004}
and the knee found in the model; this can be explained both in terms of inhomogeneity of the system and
 also in tiny variations in the parameters of the chemical evolution model, 
definitely inside the uncertainties of our modelling.

\section{Discussion}
\label{sec:disc}

\noi In this paper, we presented a new data set of chemical abundances for the
Ursa Minor (UMi) dSph, based on spectra taken with the Keck HIRES spectrograph, as well as 
a new set of chemical evolution models for 12 elements. A data set combining our new measurements with  data from  previous studies 
is consistent with a scenario in which the very metal poor stars in the dSphs have 
$\alpha$-element abundances generally comparable to those in the Galactic halo, 
although in UMi, there is a gradual decrease in the $\alpha$ elements with 
increasing metallicity which indicates that the ISM was also enriched by SN~Ia. The halo stars do not show
this trend which is a signature that the star formation in the halo
stopped on a shorter timescale compared to the timescale for
 the enrichment by SN~Ia.

\noi Our chemical evolution models use similar nucleoynthesis prescriptions to those 
by \cite{Lanfranchi2006,Lanfranchi2008} for modeling UMi with the smaller 
data set available at the time. However, our implementation of the star formation
rate is different as we derive the SFH exclusively from the CMD,
as done by \cite{North2012} for computing the manganese abundances in four other dSphs. 
Here, we extended these latter models to study Ursa Minor for the first time and to include the other 
elements for which we have measured abundances. The most striking novelty in the 
models is the nucleosynthesis adopted for
the neutron capture elements; due to the lack of data, the signature of increasing [Y/Ba] ratio 
toward lower metallicities was not studied in these previous works.
The overabundance of [Y/Ba] at very low metallicity, similar to that found in
the Galactic halo, is reproduced by  nucleosynthesis in the spinstars as 
argued in \cite{Cescutti2014} and is modeled for the first time for a dSph. This trend is due to the timescale of the onset of
r-process production by electron capture in SN~\citep[see][]{Cescutti2014};  before this point
the production of the neutron capture elements could be sustained
by s-process production in fast rotating stars \citep{Cescutti2013}.

\noi In the ratio of the [$\alpha$/Fe] abundances, it is not easy
to identify the trends in the data, ie. whether these elements show
plateaux at low metallicities up to around [Fe/H]$=-$2~dex, or whether 
they are matched better by a gradual decrease of [$\alpha$/Fe] starting from very low metallicities as found in our models. 
In either scenario, together with subsolar values of  [Ba/Fe] up to the same
metallicity where it reaches the solar value, the abundances are
consistent with a SF process that lasted a relatively long time
($\approx$ 5 Gyrs) but had a very low efficiency until it stopped at
[Fe$/$H]$=$-1.3~dex. While earlier studies found that an inefficient SF in a system without infall or mass loss seemed to fit the
data~\citep{Ikuta2002}, as the spectroscopic samples grew, already
\cite{Kirby2013} were able to show that to obtain the MDF, a model with the
infall of gas was required. Our best model is one where a gas loss
proportional to the amount of total gas was also included (winds),
and could account for both the low efficiency of the SF at later stages and
the observed lack of gas today. We emphasize that, realistically, the gas loss from the dSph
 is due to a combination of supernova feedback, winds and
tidal interactions and hence would have a more drastic effect.

\noi Our results are consistent with the previous data except for a
deviation of the abundances of the Fe-peak elements for UMI446 which is
our most metal poor star. The [Fe/H] of this star is very
similar to COS4 from \cite{Sadakane2004} which has [Fe/H]$=-$2.7 dex. 
Although abundances of the low Z $\alpha$-elements are consistent with the
rest of the sample, UMI446 has an unexpectedly higher abundance of Sc and 
the $\alpha$-elements Ti and the Fe-peak elements Cr, Mn, Ni, Cu, Zn. At Z=30 
it becomes consistent again with the rest of the data, which is especially important in the
relation of the [Y/Ba] ratio with metallicity. Among these, one of the
most challenging features is the overabundance of Mn for UMI446. These
intriguing results could point to an inhomogeneous chemical enrichment of the ISM at the early
stages. Although this inhomogeneity would be short lived,
it might be able to produce a star like UMI446 if the explosive nucleosynthesis of a SN II through a very massive progenitor with bipolar explosion (jet-like) is taken into account, as suggested by ~\cite{Maeda2003}.

\noi In the future, we plan to investigate the production of the spread in the observed abundances in UMi using more sophisticated, inhomogeneous models to have a clearer insight to its chemical evolution.

\section{Acknowledgments}
\noi  The data presented herein were obtained at the W.M. Keck Observatory, which is operated as a scientific partnership among the California Institute of Technology, the University of California and the National Aeronautics and Space Administration. The Observatory was made possible by the generous financial support of the W.M. Keck Foundation. MIW acknowledges the Royal Society for support via a University Research Fellowship. AK thanks the Deutsche Forschungsgemeinschaft for funding from  Emmy-Noether grant Ko 4161/1.


\begin{thebibliography}{}

\bibitem[Asplund et al.(2009)]{Asplund2009} Asplund, M., Grevesse, N., Sauval, A.~J. et al., ARA\&A, 47, 481

\bibitem[Ben\'itez Llambay et al.(2014)]{Benitez2014} Ben\'itez Llambay, A., Navarr J.~F., Abadi, M.~G. et al., 2014, 	arXiv:1405.5540. 

\bibitem[Bergemann \& Gehren(2008)]{Bergemann2008} Bergemann, M., Gehren, T., 2008, A\&A, 492, 823

\bibitem[Bergemann \& Cescutti(2010)]{Bergemann2010} Bergemann, M., Cescutti, G. , 2010, A\&A, 522, 9

\bibitem[Carrera et al.(2002)]{Carrera2002} Carrera, R., Aparicio, A., Martinez-Delgado, D. et al., 2012, AJ, 123, 3199

\bibitem[Castelli  \& Kurucz (2003)]{Castelli2003} Castelli, F., Kurucz, R.~L, arXiv:astro-ph/0405087

\bibitem[Cayrel et al.(2004)]{Cayrel2004} Cayrel, R., Depagne, E., Spite, M. et al., 2004, A\& A, 416, 1117 

\bibitem[Cescutti et al.(2008)]{Cescutti2008} Cescutti, G.,Matteucci, F., Lanfranchi, G.~A. et al., 2008, A \& A, 491, 401

\bibitem[Cescutti et al.(2013)]{Cescutti2013} Cescutti, G. , Chiappini, C., Hirschi, R. et al., 2013, A\&A, 553, 51

\bibitem[Cescutti \& Chiappini(2014)]{Cescutti2014} Cescutti, G., Chiappini, C., 2014, A\&A, 565, 51

\bibitem[Cohen \& Huang(2010)]{Cohen2010} Cohen J.~G., Juang, W., 2010,  ApJ,  719,  931 

\bibitem[de Boer et al.(2012)]{deBoer2012} de Boer, T.~J.~L., Tolstoy, E., Hill, V. et al., 2012, A\& A, 544, 73

\bibitem[Fran\c{c}ois et al.(2004)]{Francois2004} Fran\c{c}ois, P., Matteucci, F., Cayrel, R., Spite, M., Spite, F. et al.,  2004, A\& A, 421, 613

\bibitem[Fulbright(2000)]{Fulbright2000} Fulbright, J.~P., 2000, AJ, 120, 1841

\bibitem[Helmi et al.(2006)]{Helmi2006} Helmi, A., Irwin, M.~J., Tolstoy, E. et al., 2006, ApJ, 651, 121

\bibitem[Hendricks et al.(2014)]{Hendricks2014} Hendricks, B., Koch, A., Lanfranchi, G., 2014, ApJ, 785, 102

\bibitem[Ikuta \& Arimoto (2002)]{Ikuta2002} Ikuta, C., Arimoto, N. 2002, A\&A, 391, 55

\bibitem[Johnson(2002)]{Johnson2002} Johnson, J., 2002, ApjS, 139, 219

\bibitem[Kirby et al.(2011)]{Kirby2011} Kirby E.~N., Cohen J.~G.,  Lanfranchi, G.~A. et al.,  2011,  ApJ,  727, 78

\bibitem[Kirby \& Cohen(2012)]{Kirby2012} Kirby E.~N., Cohen J.~G.,  2012,  AJ,  144, 168 

\bibitem[Kirby et al.(2013)]{Kirby2013} Kirby E.~N., Cohen J.~G., Guhathakurta, P. et al.,  2013,  ApJ,  779, 102

\bibitem[Kleyna et al.(1998)]{Kleyna1998} Kleyna, J.~T., Geller, M.~J.,Kenyon, S.~J. et al., 1998, AJ, 115, 2359

\bibitem[Koch et al.(2006)]{Koch2006} Koch, A., Grebel, E.~K., Wyse, R.~F.~G. et al., 2006, AJ, 131, 1405

\bibitem[Koch et al.(2008)]{Koch2008} Koch, A., McWilliam, A., Grebel, E.~K. et al., 2008, ApJ, 688, 13

\bibitem[Koch et al.(2009)]{Koch2009} Koch, A., C\^ôt\'e, P., McWilliam, A., 2009, A \& A, 506, 729 

\bibitem[Koch et al.(2013)]{Koch2013} Koch, A., Feltzing, S., Ad\'en, D., Matteucci, F., 2013, A \& A, 554, 5

\bibitem[Lanfranchi et al.(2006)]{Lanfranchi2006} Lanfranchi, G.~A., Mateucci, F., Cescutti, G., 2006, MNRAS, 365, 477

\bibitem[Lanfranchi et al.(2008)]{Lanfranchi2008} Lanfranchi, G.~A., Mateucci, F., Cescutti, G., 2006,  A \& A, 481, 635

\bibitem[Maeda \& Nomoto(2003)]{Maeda2003} Maeda, K., Nomoto, K., 2003, ApJ, 598, 1163

\bibitem[Martin et al.(2007)]{Martin2007} Martin, N.~F., Ibata, R.~A., Chapman, S.~C. et al., 2007, MNRAS, 380, 281

\bibitem[McWilliam et al.(1995)]{McWilliam1995} McWilliam, A., Preston, G.,W., Sneden, C., Searle, L., 1995, AJ, 109, 2757

\bibitem[McWilliam et al.(2013)]{McWilliam2013} McWilliam, A., Wallerstein, G., Mottini, M., 2013, ApJ, 778, 149

\bibitem[Moore et al.(1966)]{Moore1966}  Moore, C.~E., Minnaert, M.~G.~J., Houtgast, J., 1966, NBS, Mono, 61

\bibitem[North et al.(2012)]{North2012} North, P., Cescutti, G., Jablonka, P. et al., 2012, A\&A, 541 , 45

\bibitem[Ram\'irez \& Mel\'endez(2005)]{Ramirez2005} Ram\'irez, I., Mel\'endez, J., 2005, ApJ, 626, 465

\bibitem[Sadakane et al.(2004)]{Sadakane2004} Sadakane K., Arimoto N., Ikuta C., Aoki W. et al., 2004, PASJ, 56, 1041  

\bibitem[Salpeter(1955)]{Salpeter1955} Salpeter, E., ApJ, 1955, 121, 161

\bibitem[Schlegel et al.(1998)]{Schlegel1998} Schlegel, D.~J., Finkbeiner, D.~P.,Davis, M., 1998, ApJ, 500, 525

\bibitem[Shetrone et al.(2001)]{Shetrone2001} Shetrone, M.~D., C\^{o}t\'e, P., Sargent, W.~L.~W., 2001, ApJ, 548, 592

\bibitem[Simon et al.(2010)]{Simon2010} Simon, J.~D., Frebel, A., McWilliam, A. et al., 2010, ApJ, 716, 446

\bibitem[Sneden(1973)]{Sneden1973} Sneden, C.~A. 1973, Ph.D. Thesis, The University of Texas at Austin

\bibitem[Stetson et al.(1998)]{Stetson1998} Stetson, P.~B., Hesser, J.~E., Smecker-Hane, T.~A. 1998, PASP, 110, 533

\bibitem[Tolstoy et al.(2004)]{Tolstoy2004} Tolstoy, E., Irwin, M.~J., Helmi, A. et al., 2008, ApJ, 617, 119

\bibitem[Tolstoy et al.(2009)]{Tolstoy2009} Tolstoy, E., Hill, V., Tosi, M., 2009, ARA\& A47, 371

\bibitem[Vogt et al.(1994)]{Vogt1994} Vogt, S.~S., Allen, S.~L., Bigelow, B.~C., 1994, SPIE, 2198, 362

\bibitem[Walker et al.(2009a)]{Walker2009a} Walker, Matthew G., Mateo, M., Olszewski, E.~W., 2009, AJ, 137, 3100 
 
\bibitem[Wilkinson et al.(2004)]{Wilkinson2004} Wilkinson, M.~I., Kleyna, J.~T., Evans, N.~W., 2004, ApJ, 611, 21.

\end{thebibliography}
\end{document}